\documentclass[reprint,nofootinbib,onecolumn,preprintnumbers]{revtex4}
\usepackage{amssymb}
\usepackage{color}
\usepackage{amsthm}
\usepackage{graphicx}
\usepackage{dcolumn}
\usepackage{bm}
\usepackage{subfigure}
\usepackage{amssymb}
\usepackage{verbatim}
\usepackage{amscd}
\usepackage{amssymb}
\usepackage{amsmath, amsfonts}
\usepackage{setspace}
\usepackage[]{graphicx}        
\usepackage{amsthm}
\usepackage{bbold}
\usepackage{enumerate}
\theoremstyle{plain}

\begin{document}

\title{Holographic State Complexity from Group Cohomology}
\author{Bart{\l}omiej Czech}
\affiliation{Institute for Advanced Study, Tsinghua University, Beijing 100084, China}


\begin{abstract}
\noindent
Topological phases of matter are often described using auxiliary systems in one extra dimension. I review the one-dimensional cluster state---the simplest quantum state with Symmetry-Protected Topological (SPT) order---as a toy model of holographic duality, and inspect it for clues about defining holographic complexity. Group cohomology, which classifies SPT orders, is a viable candidate for a robust definition of complexity in gauge/gravity duality.
\end{abstract}

\maketitle
\noindent
The concept of state complexity has taken research in the AdS/CFT correspondence by storm. Qualitatively, state complexity is understood as a CFT quantity, which is dual to the `size' of the bulk AdS spacetime. Yet despite much effort, or perhaps because of it, there is no consensus definition of complexity on either side of the duality. In the bulk, `sizes' of spacetimes can be quantified in multiple ways. The original and most popular options are the maximal volume of a spatial slice \cite{volumeprop} and the action in the Wheeler-de Witt patch \cite{actionprop1, actionprop2}, but recent work has revealed infinitely many other possibilities \cite{anythingprop}. On the boundary, there is not even a unique formalism to define state complexity. The most common approach is circuit complexity \cite{circuit1, circuit2, circuit3, circuit4}, which counts (with some cost function) unitary gates necessary to convert a designated (usually unentangled) reference state into the desired state. But there are multiple alternatives, including path integral optimization \cite{pio1, pio2, pio3}, variants of information-geometric distances \cite{compldistance1, compldistance2}, and others. Even if we accept circuit complexity as a preeminent paradigm, its inherent ambiguities pose a further question \cite{complrg}: how to classify gate costs into universality classes according to their infrared (large complexity) behaviors?

The many unresolved differences and ambiguities involved in defining complexity call for a principled, rules-based approach. By what criteria should we judge a proposal for holographic complexity? I submit the following {\bf desiderata}:
\begin{enumerate}
\item In the CFT, complexity should involve concepts, which play salient roles in bulk reconstruction. A nonexhaustive list of such concepts includes reduced states \cite{rt1, rt2, hrt, subregion} and modular Hamiltonians \cite{flm, jlms}, local bulk operators \cite{hkll, subregionproof, subregionchannel}, modular Berry connections and curvatures \cite{mbp, mpt, modularchaos}, and others \cite{ewcs, mme, symplecticform, propertime}. If we do not start with holographically meaningful ingredients, obtaining a holographically meaningful answer will be a miracle.
\item A definition of complexity should be faithful to the spirit of the original motivation for relating bulk `sizes' to state complexity \cite{volumeprop, tomjuan}. Probably the best illustration of the original thinking is a tensor network: if counting tensors in a network enumerates steps necessary to prepare the state, and if a tensor network models a bulk spacetime, then state complexity should roughly go like bulk volume. 
\item As in all physical modeling, a proposal is stronger if it involves fewer free parameters. In an ideal proposal, all parameters would be internally or dynamically determined.  
\item Ideally, holographic complexity should mesh with some broader conceptual framework, which extends beyond AdS/CFT. For this desideratum, I see two main possibilities: One is bulk/boundary duality that is not gauge/gravity duality, often used for describing topological phases of matter \cite{frs, stringnet, xgwspt, xgwrecent}. Another one is complexity theory as studied by computer scientists. Other things being equal, a proposal for holographic complexity is stronger if it connects to one or both of these broader frameworks. 
\end{enumerate}

The purpose of this talk is to illustrate how these demands may be simultaneously met. Partly to situate holographic complexity in a broader framework, and partly for pure inspiration, I first consider an entirely different system: a one-dimensional array of qubits in an entangled state with symmetry-protected topological order (SPT). I argue that this system shows a strong analogy to the AdS/CFT correspondence. Under the analogy, the `size' of the `bulk' is quantified by a group cohomology relevant to the SPT order. I then outline how group cohomology may be relevant to defining complexity in gauge/gravity duality. 

\begin{figure}
        \centering
        \includegraphics[width=0.50\textwidth]{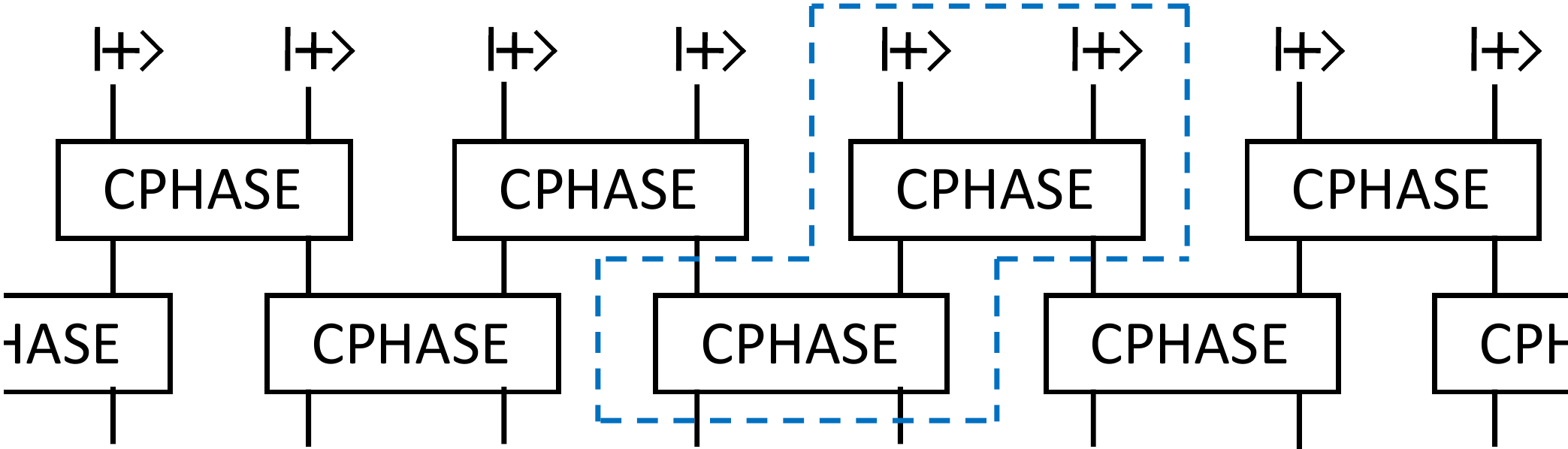}
        \hfill
        \raisebox{0.03\textwidth}{\includegraphics[width=0.19\textwidth]{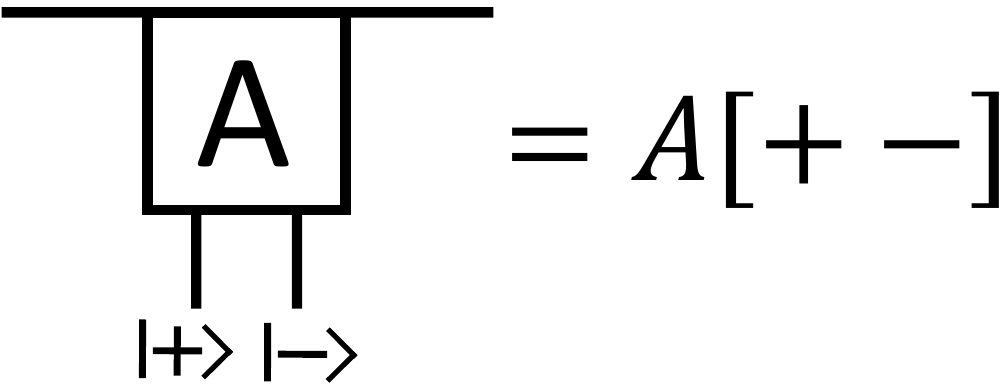}}      
        \hfill
        \includegraphics[width=0.14\textwidth]{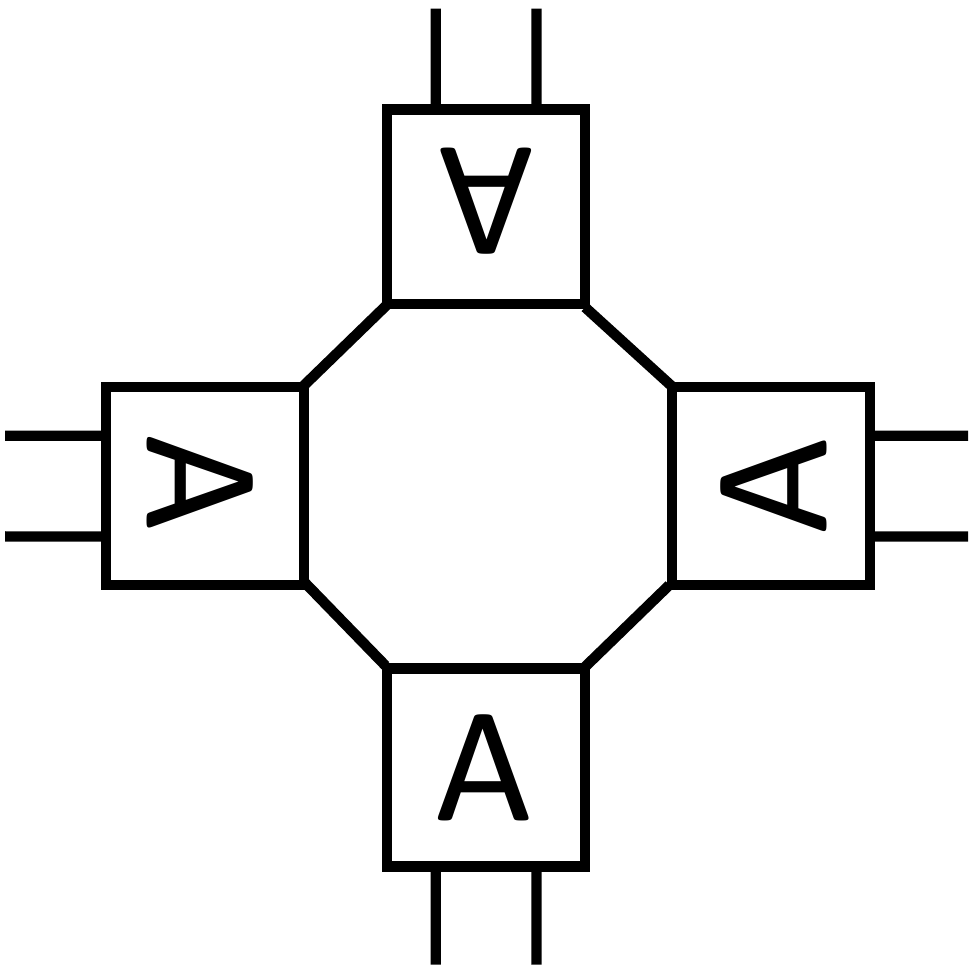}
        \caption{The cluster state, here shown for eight qubits on a loop, as given by equation~(\ref{cluster1st}) (left panel) and equation~(\ref{mpsdesc}) (right panel). The highlighted part of the left panel is the `matrix' of the MPS description (middle panel).}
        \label{fig:clusterbasic}
\end{figure}

To avoid awkward generalities and an onslaught of definitions, I focus on a particular SPT-ordered state, called a cluster state \cite{cluster}; see \cite{beni} for a pedagogical presentation. It is obtained (see Figure~\ref{fig:clusterbasic}) by acting with entangling gates 
\begin{equation}
{\rm CPhase}^{i,i+1} = {\rm diag}\{1,1,1,-1\}^{i,i+1} = \mathbb{1}^{i,i+1} - 2\,|11\rangle^{i,i+1} \langle 11| \qquad \textrm{(in $\sigma_z^i \otimes \sigma_z^{i+1}$ basis)}
\label{cluster1st}
\end{equation}
on each pair of neighboring spins, initialized in $+1$ eigenstates of $\sigma_x$, denoted $|+\rangle = \sigma_x |+\rangle$:
\begin{equation}
|{\rm cluster}\rangle 
= \prod_i {\rm CPhase}^{i,i+1}\, \bigotimes_i |+\rangle^i
\label{clusterproduct}
\end{equation} 
It is most convenient to consider an even number $2K$ of spins arranged on a loop, so that the product in (\ref{clusterproduct}) contains CPhase$^{2K, 2K+1} \equiv {\rm CPhase}^{2K,1}$. Then $|{\rm cluster}\rangle$ is the unique ground state of the Hamiltonian:
\begin{equation}
H = - \sum_i \sigma_z^{i} \sigma_x^{i+1} \sigma_z^{i+2}
= - \left( \prod_i {\rm CPhase}^{i,i+1} \right) \left( \sum_i \sigma_x^i \right) \left( \prod_i {\rm CPhase}^{i,i+1} \right)
\end{equation}
It follows that the state is stabilized by each operator $\sigma_x^{i} \sigma_z^{i+1} \sigma_x^{i+2}$ and by products of them. Using now that the number of spins is assumed even, we identify two special stabilizing operators: 
\begin{equation}
\prod_{i~{\rm even}} \sigma_z^{i} \otimes \sigma_x^{i+1} \otimes \sigma_z^{i+2} 
= \prod_{j~{\rm odd}} \sigma_x^j \equiv S_{\rm odd}
\qquad {\rm and} \qquad
\prod_{i~{\rm odd}} \sigma_z^{i} \otimes \sigma_x^{i+1} \otimes \sigma_z^{i+2} 
= \prod_{j~{\rm even}} \sigma_x^j \equiv S_{\rm even}
\end{equation}
They generate an on-site $\mathbb{Z}_2 \otimes \mathbb{Z}_2$ symmetry of the state, which is not respected by the entangling gate in equation~(\ref{clusterproduct}): $[{\rm CPhase}^{i,i+1}, S_{\rm odd}] \neq 0 \neq [{\rm CPhase}^{i,i+1}, S_{\rm even}]$. Thus, although it is possible to bring the cluster state to an unentangled (product) form with short range disentangling gates---as showcased by equation~(\ref{clusterproduct})---it is not possible to do so without breaking the on-site $\mathbb{Z}_2 \otimes \mathbb{Z}_2$ symmetry. This is `Symmetry-Protected Topological order' (SPT).  

I now argue that the cluster state is a toy model of holographic duality. The reasoning, which backs this assertion takes several steps. The first is to rewrite $|{\rm cluster}\rangle$ as a Matrix Product State (MPS). A little play with Pauli matrices (see Figure~\ref{fig:clusterbasic}) reveals that its amplitudes in the $|\pm\rangle = \pm \sigma_x |\pm\rangle$ basis are:
\begin{align}
\langle (s_1 s_2) \otimes (s_3 s_4) \otimes \ldots \otimes (s_{2K-1} s_{2K}) | {\rm cluster} \rangle &
= 2^{1-2K} {\rm Tr}\, A[s_1 s_2] A[s_3 s_4] \ldots A[s_{2K-1} s_{2K}]
\label{mpsdesc}
\\
{\rm with} \quad A[++] = \mathbb{1} \quad {\rm and} \quad 
A[+-] = -\sigma_z \quad {\rm and} & \quad 
A[-+] = -\sigma_x  \quad {\rm and} \quad 
A[--] = \sigma_x \sigma_z 
\label{ourprojective}
\end{align}
The $(s_i s_{i+1})$ in (\ref{mpsdesc}) range over the set $\{++, +-, -+, --\}$ and denote $\sigma_x^i \otimes \sigma_x^{i+1}$ eigenstates. The objects $A[s_i s_{i+1}]$ are the `matrices' in the matrix product description of $|{\rm cluster}\rangle$. 

\begin{figure}
        \centering
        \includegraphics[width=0.12\textwidth]{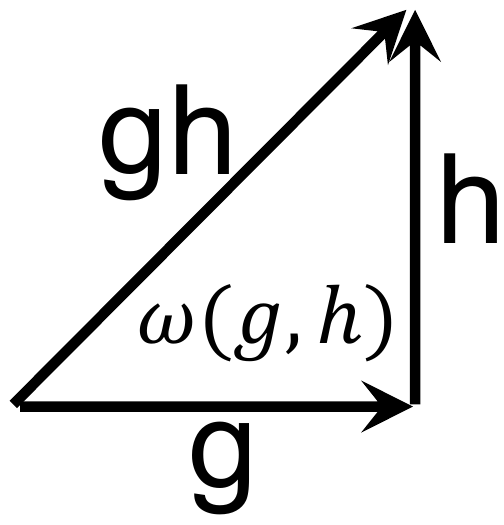}        
        \hspace*{0.2\textwidth}
        \includegraphics[width=0.35\textwidth]{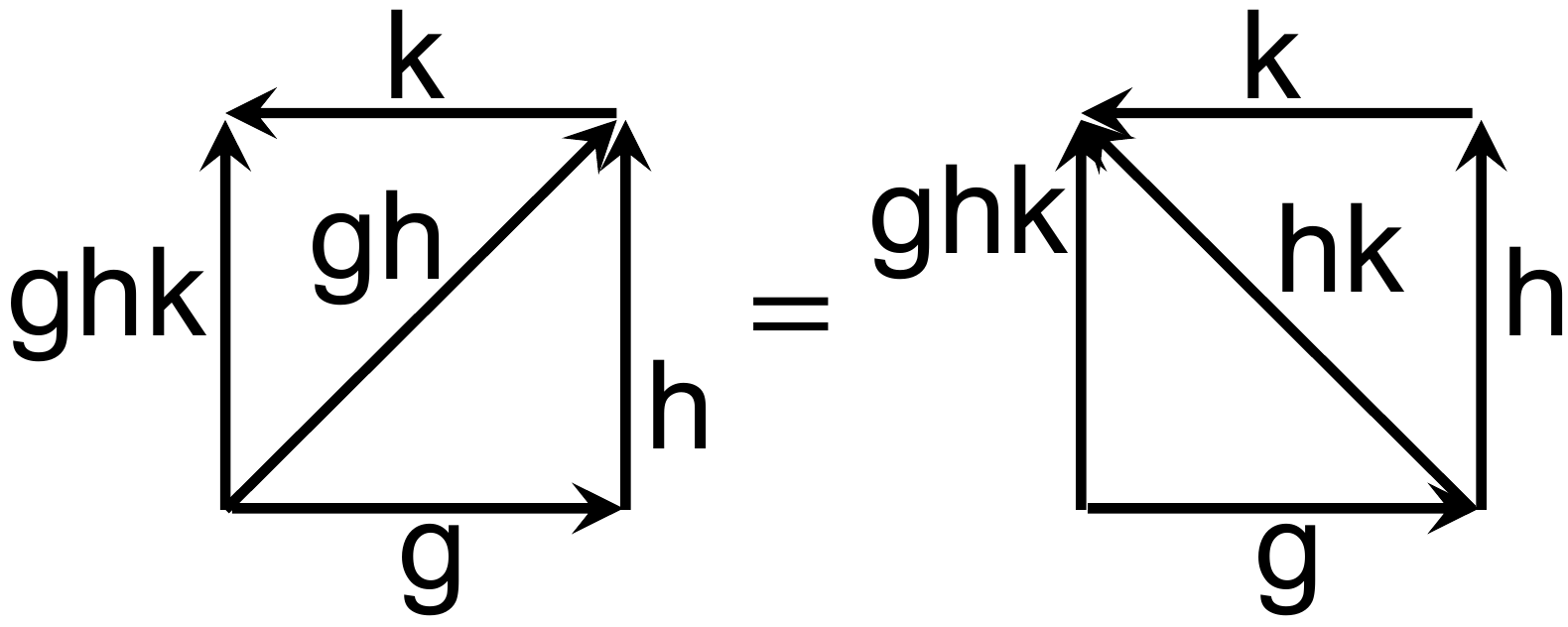}
        \caption{The factor $\omega(g,h)$ in equation~(\ref{projectiverep}) can be visualized as a triangle with oriented edges, which run neither clockwise nor counterclockwise. Associativity of the group product mandates equation~(\ref{cocycle}), visualized in the right panel. Therefore products of $\omega$'s are represented by triangulation-invariant two-complexes and $\log \omega$ behaves like a closed two-form.}
        \label{fig:simplices}
\end{figure}

The next step is to think about the $(s_i s_{i+1}) \in \{++, +-, -+, --\}$ as elements of $\mathbb{Z}_2 \times \mathbb{Z}_2$---the group that protects the topological order of $|{\rm cluster}\rangle$. Observe that $A[s_i s_{i+1}]$ form a projective representation of the group:
\begin{equation}
A[g] A[h] = \omega(g,h) A[gh] \qquad \forall g,h \in \mathbb{Z}_2 \times \mathbb{Z}_2
\label{projectiverep}
\end{equation}
An example of a nontrivial $\omega$, which makes equation~(\ref{ourprojective}) into a projective and not an ordinary (affine) representation, is $A[--] A[--] = -\mathbb{1} = -A[++]$, so $\omega(--, --) = -1$. 

It is well known that both SPT phases of matter \cite{xgwspt} and projective representations of groups are classified by the second group cohomology $H^2(G, U(1))$. For representations, this is easy to understand. First, associate $\omega(g,h)$ with an oriented two-simplex whose edges are marked (in the order fixed by the orientation) with $g \in G$, $h \in G$, and $gh \in G$; see Figure~\ref{fig:simplices}. If the representation matrix $A[(gh)k] = A[g(hk)]$ is to be well-defined, we must have:
\begin{equation}
\omega(g,h) \omega(gh, k) = \omega(g,hk) \omega(h,k)
\label{cocycle}
\end{equation}
This means that $\log \omega(g,h)$ behaves like a closed two-form or---figuratively speaking---like the `area' of a `triangle' with sides $g,h, gh$. If we now choose to dress the MPS matrices $A[g]$ with $g$-dependent phases, $A[g] \to \mu(g) A[g]$, then $\omega(g,h) \to \omega(g,h) \mu(g) \mu(h) / \mu(gh)$. Applied to equation~(\ref{mpsdesc}), this operation dresses $|{\rm cluster}\rangle$ with an irrelevant overall phase, so we would like to treat such $\omega$'s as equivalent. In group cohomology, this redefinition is understood as adding to the `two-form' $\log \omega(g,h)$ the exterior derivative of a `one-form' $\log \mu(g)$. 

\begin{figure}
        \centering
        \hspace*{0.1\textwidth}
        \includegraphics[width=0.19\textwidth]{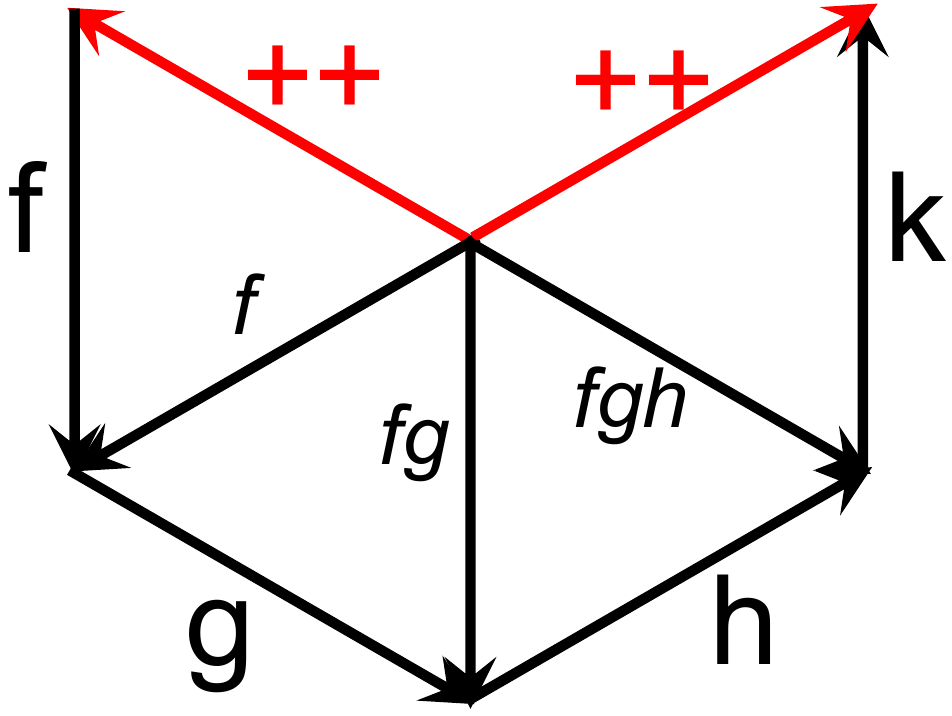}
        \hfill
        \includegraphics[width=0.19\textwidth]{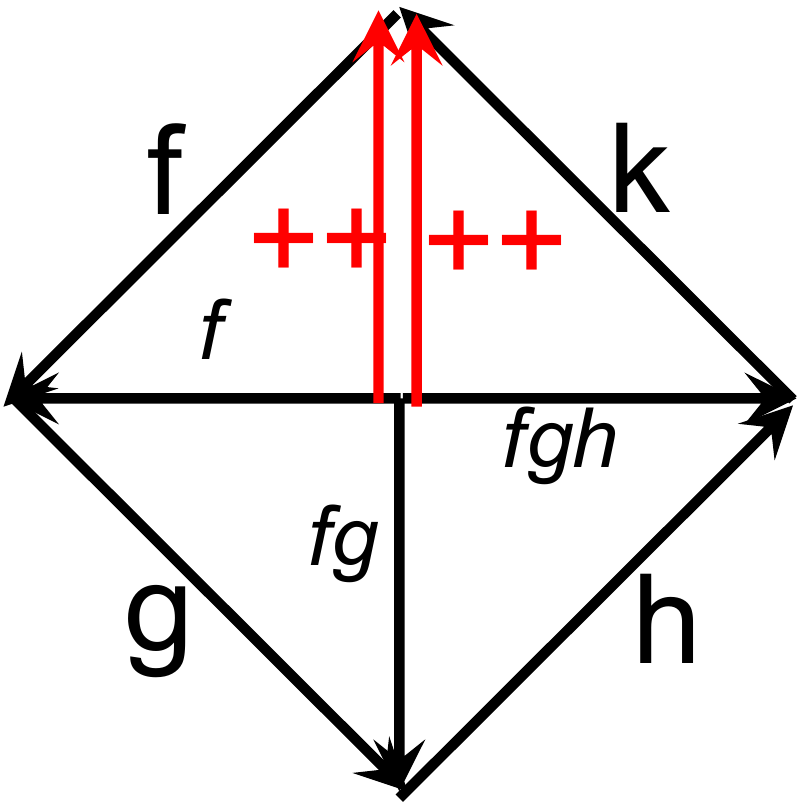}
        \hfill
        \includegraphics[width=0.19\textwidth]{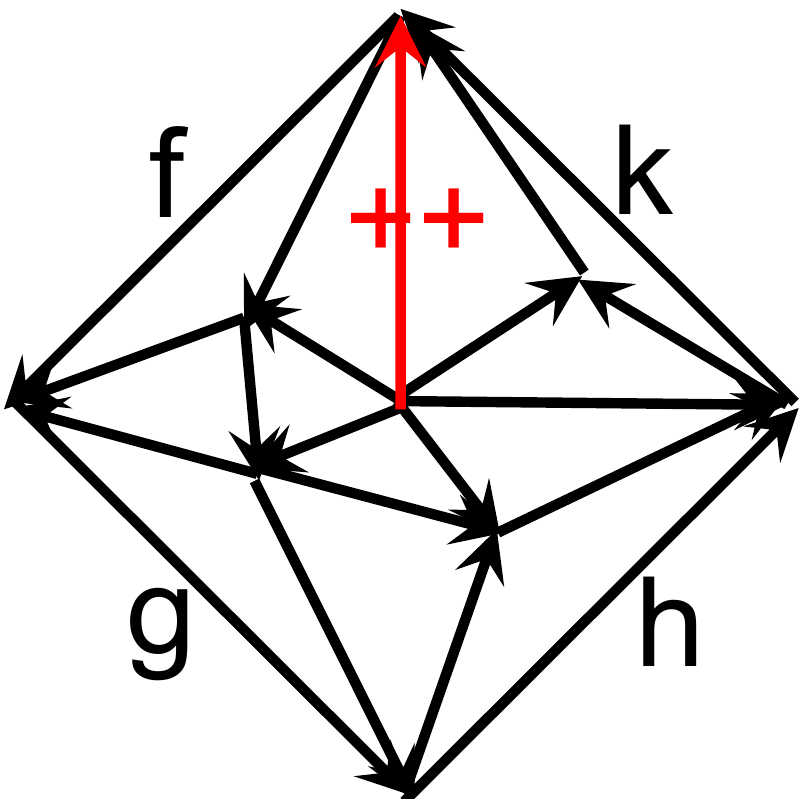}
                \hspace*{0.1\textwidth}
        \caption{The cluster state on eight qubits (also shown in Figure~\ref{fig:clusterbasic}) is prepared by the path integral of a discrete two-dimensional topological field theory with action $S = i \log \omega(g,h)$. I label the basis $|f\rangle\, |g\rangle\, |h\rangle\, |k\rangle$ used in equation~(\ref{tftpi}). Interior edges, which are summed in the path integral, are marked in italics whenever they are uniquely determined. All three pictures are equivalent because of triangulation invariance (Figure~\ref{fig:simplices}). The progression from the left to the right panel shows that the boundary condition $|\!+\!+\rangle \langle\!+\!+|$, which sets $k = (fgh)^{-1}$ and imposes ${\rm Tr}(\ldots)$ in equation~(\ref{mpsdesc}), is contractible in the sense that the path integral is effectively carried out over the half-sphere without punctures.}
        \label{fig:clustertft}
\end{figure}

We are now ready to see the `holographic' description of the cluster state. For this, substitute equation~(\ref{projectiverep}) in (\ref{mpsdesc}) and use the graphical representation of $\omega(g,h)$. For example, the circular cluster on $2K = 8$ qubits can be written as:
\begin{equation}
|{\rm cluster}\rangle = \frac{1}{64}
\sum_{f,g,h \in \mathbb{Z}_2 \times \mathbb{Z}_2} 
\left\{
\omega(++,f) \, \omega(f, g)\, \omega(fg, h)\, 
\omega\big(fgh, (fgh)^{-1} \big) 
\right\}
\,\,|f\rangle\, |g\rangle\, |h\rangle\, |(fgh)^{-1}\rangle\,,
\label{tftpi}
\end{equation}
where again we use the notation $|f\rangle$ with $f \in \mathbb{Z}_2 \times \mathbb{Z}_2$ for $\sigma_x^i \otimes \sigma_x^{i+1}$ eigenstates $|\!+\!+\rangle$, $|\!+\!-\rangle$, $|\!-\!+\rangle$, and $|\!-\!-\rangle$. This state is shown in Figure~\ref{fig:clustertft}. 

Using equation~(\ref{cocycle}) and Figure~\ref{fig:simplices}, we see that the graphical representations of $|{\rm cluster}\rangle$ in Figure~\ref{fig:clustertft} are triangulation-invariant. In the spirit of holography, we interpret the complexes of triangles as path integrals of a bulk theory, which is tasked with preparing $|{\rm cluster}\rangle$. The action of the bulk theory on a triangle with sides $g, h$ and $gh$ is given by $e^{-iS} = \omega(g,h)$ if the $g,h$-sides make a counterclockwise pattern, and $e^{+iS} = \omega(g,h)$ otherwise. On interior edges, the correctly normalized measure of the bulk path integral is $\mathcal{D}g = (1/4) \sum_{g \in \mathbb{Z}_2 \times \mathbb{Z}_2}$. 

This bulk theory is a discrete topological field theory (TFT), where `topological' connotes triangulation invariance. Every discrete SPT-ordered system can be prepared by a path integral of this type of TFT \cite{xgwspt}. An MPS description of SPT phases \cite{ryu}, which we used in equation~(\ref{mpsdesc}), was developed in generality based on the bulk TFT perspective. Continuum versions of these bulk theories, called equivariant topological field theories, are also well-studied \cite{dw}. 

\begin{figure}
        \centering
        \includegraphics[width=0.19\textwidth]{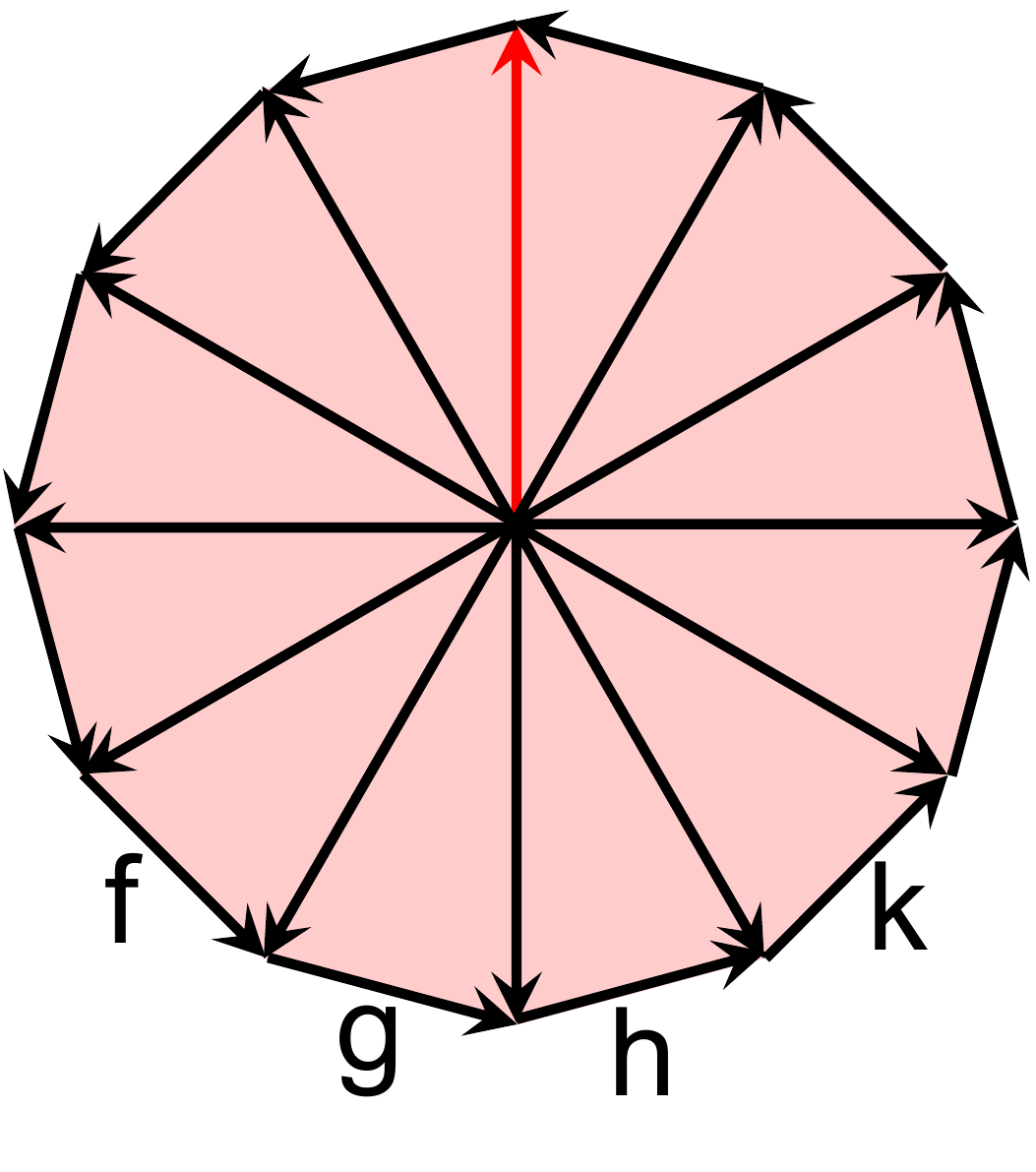}
        \hspace*{0.05\textwidth}
        \includegraphics[width=0.19\textwidth]{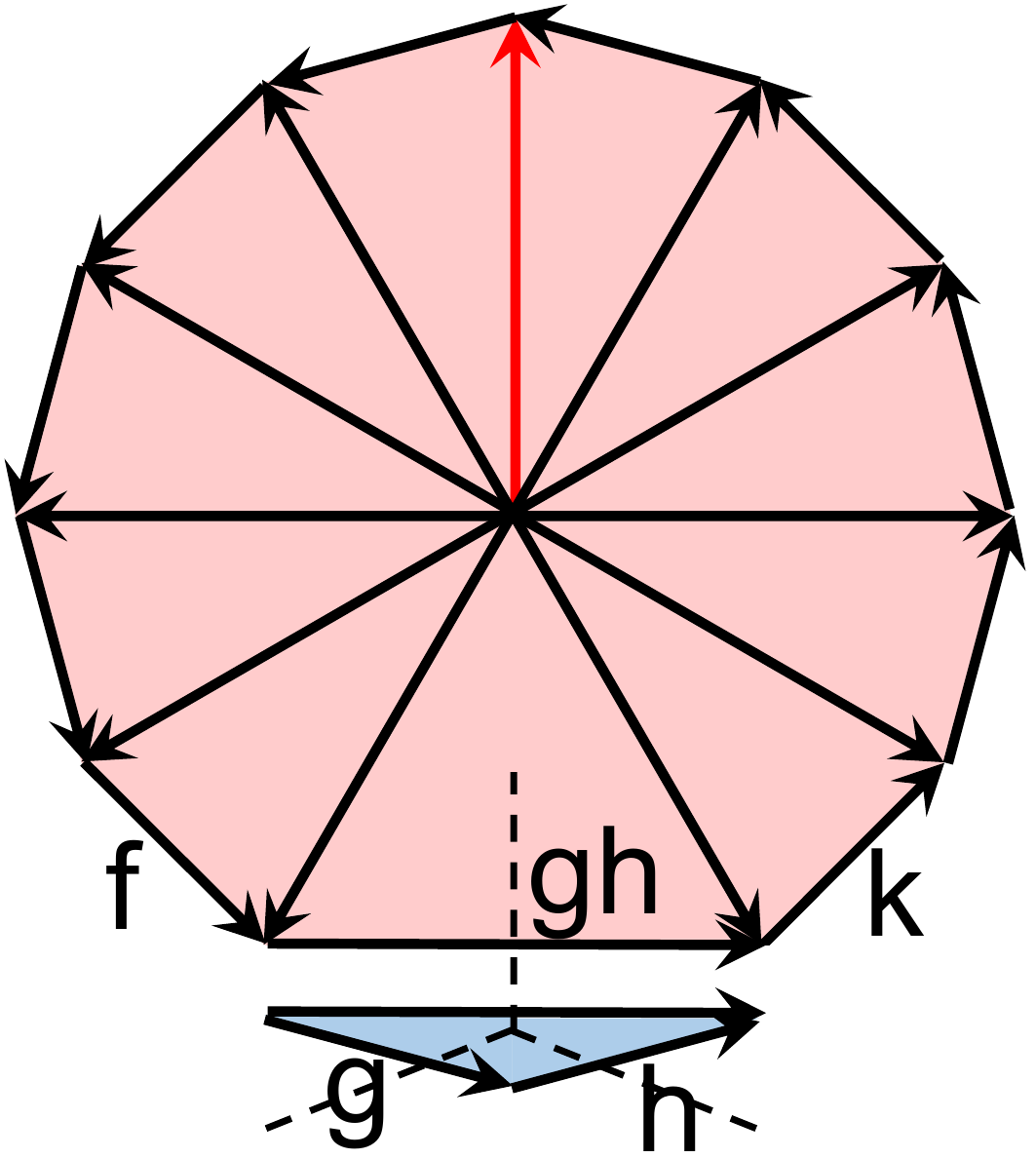}
        \hspace*{0.15\textwidth}
        \includegraphics[width=0.25\textwidth]{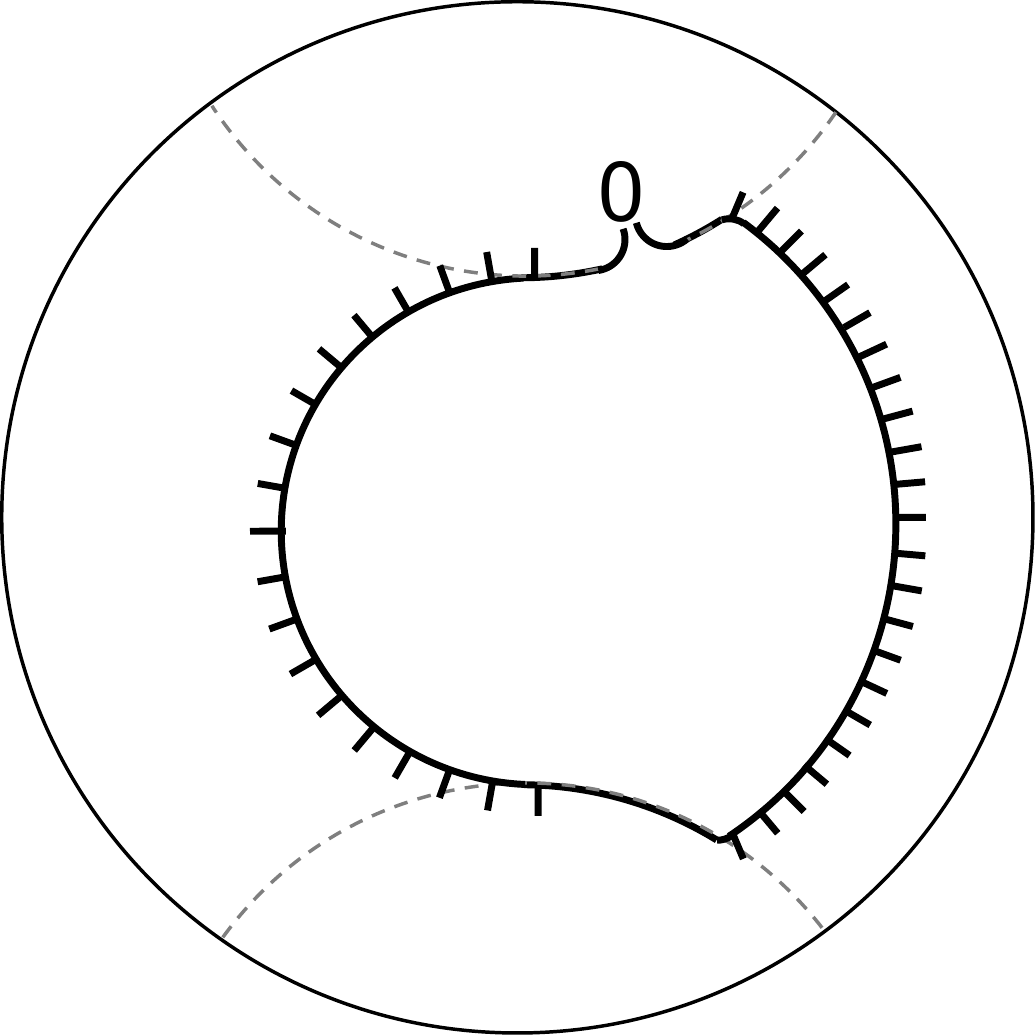}
        \caption{Using triangulation invariance (Figure~\ref{fig:simplices}), we can split the cluster state on 24 qubits (`ultraviolet,' left panel) into a cluster on 22 qubits (`infrared,' middle panel) and a gate, which fuses four `ultraviolet' qubits in state $|g\rangle |h\rangle$ into two `infrared' qubits in state $|gh\rangle$. This is a local application of holographic Renormalization Group in SPT holography. The right panel shows an independently derived \cite{wilsonnetwork} MPS form of the vacuum of a holographic CFT$_2$ at a variable UV cutoff, which is an amputated Wilson line network of $SL(2,\mathbb{R})\times SL(2,\mathbb{R})$ Chern-Simons theory. Observe the similarity to the cluster state, with the $|0\rangle\langle 0|$ insertion functioning like $|\!+\!+\rangle\langle+\!+\!|$ in Figure~\ref{fig:clustertft}.}
        \label{fig:states}
\end{figure}

I now highlight features of SPT holography, which are analogous to gauge/gravity duality:
\begin{itemize}
\item Gauge/gravity duality is often modeled with tensor networks such as MERA \cite{mera} or HaPPY \cite{happy}. The preparation of $|{\rm cluster}\rangle$ in equation~(\ref{clusterproduct}) is like a single layer of those networks; see the left panel of Figure~\ref{fig:clusterbasic}. There is only a single layer because SPT phases have only short range entanglement \cite{xgwspt}. 
\item The MPS presentation of $|{\rm cluster}\rangle$ involves a projective representation of the symmetry group. This is analogous to the gauge/gravity lore that global boundary symmetries are local bulk symmetries.
\item The triangulation invariance of the TFT path integral is analogous to diffeomorphism invariance in gravity.
\item Triangulation invariance also motivates a notion of local Renormalization Group analogous to holographic RG of gauge/gravity duality; see Fig.~\ref{fig:states}. The same triangle can be considered part of the preparation of the (ultraviolet) state, or excluded from the preparation of the (infrared) state and absorbed by a redefinition of the cutoff.  
\item As a slight generalization of $|{\rm cluster}\rangle$, we can consider states \cite{ryu} whose amplitudes are as in equation~(\ref{mpsdesc}), but with the replacement ${\rm Tr}\,(\ldots) \to {\rm Tr}\,(\Sigma \ldots)$ where $\Sigma = \sigma_x$, $\sigma_z$, or $\sigma_x \sigma_z$. Their path integral preparations, as shown in Figure~\ref{fig:clustertft}, would have one $++$ replaced with $+-$, $-+$, or $--$, so the rightmost panel of Figure~\ref{fig:clustertft} would contain a defect-carrying bulk puncture. This is like a black hole in gauge/gravity duality. In SPT holographic duality, the product $fghk \in G$ equals the $G$-valued `charge' of the defect; this is like Gauss's law in gravity.  
\end{itemize}

In this analogy, there is an obvious quantity, which satisfies points 2.-4. from our list of desiderata for holographic complexity. It is the action of the bulk TFT. Qualitatively speaking, it `counts' the gates necessary to map the product state $\otimes_i |+\rangle^i$ to $|{\rm cluster}\rangle$ in equation~(\ref{clusterproduct}). But it does so in an unambiguous way because the complexity measure is constrained by the topological (triangulation) invariance of the bulk. Indeed, the action of the bulk is essentially the only sensible measure of complexity applicable to topological field theories \cite{circuittft}. At the same time, the TFT action is very much like the bulk spatial volume in gauge/gravity duality because---as I emphasize in equation~(\ref{cocycle}) and in Figure~\ref{fig:simplices}---the `Lagrangian density' $i \log \omega$ behaves like a closed two-form. In this sense, the bulk action is the closest possible analogue for quantifying bulk sizes, which makes sense in a topological field theory. 

The parallel between SPT phases and gauge/gravity duality validates the intuition behind the volume \cite{volumeprop} and action \cite{actionprop1} proposals. It does not, and cannot, lend them quantitative support because in SPT holography we have a $(d+1)$-dimensional action while the gauge/gravity proposals call for a $(d+2)$-dimensional action or a $(d+1)$-dimensional volume---which the TFT cannot see. A more compelling match is with path integral optimization when it ends up identifying constant mean-curvature slices of the bulk geometry \cite{pio4, pio6}.

But a more daring lesson that might be drawn from SPT phases concerns complexity and group cohomology. For the cluster state, a cocycle of its on-site symmetry group supplies a measure of complexity, which captures the `size' of the bulk description. Can the same be done in gauge/gravity duality? It seems that the answer is yes. 

In using group cohomology to define a bulk density of complexity, we must have an on-site symmetry $G$ acting at every location on the bulk cutoff surface. The boundary degrees of freedom must transform in an ordinary (affine) representation of $G$ while the bulk degrees of freedom must transform projectively. We should also have a flat $G$-connection in the bulk; see Figures~\ref{fig:simplices} and \ref{fig:clustertft}. The group cohomology $H^{d+1}(G, M)$, which is relevant to describing states in $d$ spatial boundary dimensions, may have $M = \mathbb{R}$ or $M = U(1)$. Then (logarithms of) cocycles drawn from $H^{d+1}(G, M)$ automatically give well-defined bulk densities because they are essentially closed $(d+1)$-forms.

These prerequisites are easily met by gauge/gravity duality. In fact, there is already an explicit example: the complexity of the ground state of a holographic CFT$_2$ at a variable UV cutoff, which my collaborators and I studied in \cite{wilsonnetwork}. There, we presented the ground state in an MPS-like form (see Figure~\ref{fig:states}) as an amputated network of Wilson lines of the $SL(2,\mathbb{R})\times SL(2,\mathbb{R})$ Chern-Simons theory, whose flat solutions describe locally AdS$_3$ spacetimes \cite{ads3cs}. We then argued that the density of complexity in this setup should go like bulk area. But $e^{-A_\triangle}$, where $A_\triangle$ is the area of a triangle in hyperbolic space, is a cocycle of $SL(2,\mathbb{R})$---a fact we instantly recognize by looking back at Figure~\ref{fig:simplices}.

Can group cohomology define a measure of complexity more broadly, beyond locally AdS$_3$ spaces? The key is to identify a group, which acts as an on-site symmetry everywhere on the cutoff surface. In \cite{wilsonnetwork}, this is $SL(2,\mathbb{R})$---the global isometry group of two-dimensional hyperbolic space. In a general context, where the bulk has no global isometries, the only geometric (not internal) on-site symmetry available is the \emph{local} isometry group, i.e. the Poincar{\'e} group. In the bulk, its cohomology gives rise to measures of complexity, which effectively produce integrals of bulk curvature invariants---and so agree well with the heuristics of holographic complexity. In the boundary language, generators of the bulk Poincar{\'e} symmetry are partly understood in terms of modular scrambling modes \cite{modularchaos}, modular zero modes including the modular Hamiltonian \cite{jlms}, and a subtle combination thereof in the case of bulk time translations \cite{propertime}. Assuming this understanding can be advanced, defining holographic complexity from the cohomology $H^{d+1}(\textrm{Poincar{\'e}}, \mathbb{R})$ will have met desideratum 1.

Before closing, I briefly revisit desiderata 3. and 4. First, group cohomology is a tight algebraic structure. While it cannot fix a definition of holographic complexity uniquely, it imposes on it a set of stringent rules and eliminates spurious ambiguities. The second point concerns embedding holographic complexity in a broader framework. SPT phases in general, and the cluster state in particular, are resource states for Measurement-Based Quantum Computation \cite{mbqc}---a framework for doing quantum computing by measurements alone. There is evidence that group cohomology also quantifies the computer scientist's complexity of certain algorithms carried out in the MBQC or MBQC-like framework \cite{wilsonnetwork, inprep}. If correct, this would be a remarkable reunification of the term `complexity' with its original, non-holographic meaning.

\paragraph*{Acknowledgments} 
\noindent
I thank Yang-Hui He for his kind invitation to present this talk at the 2021 Nankai Symposium on Mathematical Dialogues. The talk is based in part on two ongoing collaborations: one with Jan de Boer, Bowen Chen, Lampros Lamprou and Zi-zhi Wang, and one with Robert Raussendorf and Gabriel Wong. I have also benefitted greatly from conversations with Ling-Yan (Janet) Hung, as well as Jingyuan Chen and Runze Chi.

\end{document}